\documentclass{IEEEoj}
\usepackage{cite}
\usepackage{amsmath,amssymb,amsfonts}
\usepackage{graphicx,color}
\usepackage{textcomp}

\usepackage{algorithm, algorithmicx}
\usepackage{algpseudocode}
\usepackage{xcolor}
\usepackage[table]{xcolor}
\usepackage{multirow}
\usepackage{pifont}
\usepackage{subcaption}
\usepackage{graphicx}
\usepackage{booktabs}
\usepackage{url}
\usepackage{tabularx}
\usepackage{booktabs}
\usepackage{wasysym}
\usepackage{etoolbox}

\newcommand{\newtool}[1]{{5G-Shark}}
\newcommand{\cmark}{\ding{51}}%
\newcommand{\xmark}{\ding{55}}%

\usepackage[table]{xcolor}
\definecolor{lightgreen}{HTML}{E6F4EA}
\definecolor{lightred}{HTML}{FCE8E6}
\newcommand{\legendbox}[2]{%
  \bgroup
  \setlength{\fboxsep}{0pt}
  \setlength{\fboxrule}{1.2pt}
  \fcolorbox{#1}{#2}{\rule{0pt}{6pt}\rule{10pt}{0pt}}
  \egroup
}

\def\BibTeX{{\rm B\kern-.05em{\sc i\kern-.025em b}\kern-.08em
    T\kern-.1667em\lower.7ex\hbox{E}\kern-.125emX}}
\AtBeginDocument{\definecolor{ojcolor}{cmyk}{0.93,0.59,0.15,0.02}}
\def\OJlogo{}

\newcommand{\submissionnotice}{This work has been submitted to the IEEE for possible publication. Copyright may be
transferred without notice, after which this version may no longer be accessible.}
\makeatletter
\long\def\receivedfont#1\par{}%
\let\ps@plain@orig\ps@plain
\renewcommand{\ps@plain}{%
\ps@plain@orig
\def\@oddhead{\vbox{\hsize\textwidth\footnotesize\noindent\submissionnotice}}%
\let\@evenhead\@oddhead
}%
\makeatother

\begin{document}
\bstctlcite{BSTcontrol}

\long\def\receivedfont#1\par{}

\title{\newtool{}: A Network Security Auditor for 5G Subscriber Privacy and Unauthenticated Signalling Resilience}

\author{
    Oscar Lasierra\IEEEauthorrefmark{1}, 
    Gines Garcia-Aviles \IEEEauthorrefmark{1}, 
    Antonio Skarmeta \IEEEauthorrefmark{2},
    AND  Xavier Costa-Pérez \IEEEauthorrefmark{1,3,4} \IEEEmembership{(Senior Member, IEEE)}
}

\affil{i2CAT Foundation}
\affil{University of Murcia}
\affil{NEC Laboratories Europe}
\affil{ICREA}

\corresp{CORRESPONDING AUTHOR: Gines Garcia-Aviles (e-mail: gines.garcia@i2cat.net).}

\authornote{This work was supported in part by the ORIGAMI Project under Grant 101139270; in part by the CERCA Programme from the Generalitat de Catalunya through the ICREA programme; and in part by the funding received from the Department de Recerca I Universitats, Generalitat de Catalunya for this project}

\markboth{\newtool{}: A Network Security Auditor for 5G Subscriber Privacy and Unauthenticated Signalling Resilience}{Lasierra \textit{et al.}}

\begin{abstract}

The fifth generation of mobile networks (5G) was standardised with an explicit mandate to close long-standing privacy and security gaps, mandating the concealment of the subscriber's permanent identity, resistance to generational downgrade, and protection against location tracking. Assessing whether these guarantees hold in operational networks, however, requires separating two sources of residual exposure that prior studies do not distinguish and do not evaluate in the wild: \emph{protocol-design limitations}, which remain exploitable even against a fully specification-compliant deployment, and \emph{implementation gaps}, which arise from incomplete or non-compliant implementations. We present \newtool{}, a low-cost security assessment tool and methodology that turns a legitimate mobility procedure against the subscriber. Rather than relying on active jamming or malformed-packet injection, \newtool{} manipulates the standardised cell-reselection criterion to pull a target User Equipment (UE) onto a self-created rogue cell, establishing an attack vantage with minimal service disruption and a low entry barrier. Then, the proposed methodology effectively performs the required interactions to expose the security risks of the system under test, classifying them into the aforementioned categories. Built solely from open-source stacks and inexpensive Software Defined Radio (SDR) hardware and evaluated against commercial 5G Standalone (SA) deployments, \newtool{} requests subscriber identifiers, forces Radio Access Technology (RAT) downgrade via crafted \textit{Registration Reject} cause codes, and induces denial-of-service states. For each vector, we attribute the root cause to protocol design or deployment non-compliance. We further provide empirical evidence that in several commercial deployments, temporary identifiers are re-allocated in near-sequential steps that keep successive values linkable, a weakness that enables persistent user tracking despite correct subscriber ID concealment.


\end{abstract}


\maketitle

\section{INTRODUCTION} \label{sec:intro}

The fifth generation of mobile networks was standardised with an explicit mandate to close the privacy and security gaps inherited from previous generations. Among the improvements, three key innovative guarantees anchor this mandate: i) the concealment of the subscriber's permanent identity at the radio interface; ii) resistance to being downgraded to previous generations; and iii) protection against location tracking. In 4G and previous generations, permanent identifiers were exchanged over the air in clear text, making their interception relatively easy for an attacker, enabling unauthorised tracking and surveillance through International Mobile Subscriber Identity (IMSI) catching~\cite{imsi-catching-defcon}. In response, 5G introduced the concealment of the Subscription Permanent Identifier (SUPI), where, before transmitting over the air, the SUPI is encrypted into a Subscription Concealed Identifier (SUCI), which should prevent third-party observers from recovering the subscriber's permanent identity. Then, the inclusion of the Anti-Bidding Down Between Architectures (ABBA) parameter reduced the probability of forcing devices to perform downgrades to previous (and vulnerable) generations. Finally, the identification of the low randomness in the temporary identifiers generation induced the standard to propose robust strategies for their generation (higher refresh rates or more robust seed reselection), but they still remain optional in the standard.

Assessing whether these guarantees hold in practice, however, requires separating two sources of residual exposure that are usually joined. The first are \emph{protocol-design limitations}: properties that remain exploitable even against a \textit{fully specification-compliant} deployment. These vulnerabilities are properties of the specific procedure itself that require changing the standard procedure to cover their limitations. For example, a subscriber will always emit a SUCI when solicited, because it must provide an identity to register, a pre-authentication \textit{Registration Reject} is processed before any Non-Access Stratum (NAS) security context exists, or the cell-reselection criteria will be performed by the subscriber to jump into connected mode or link degradation. The second are \emph{implementation gaps}: exposures that arise from incomplete, non-compliant or simpler implementations where even though the standard offers a secure path, the exposure appears because the real deployment/modem implements a weak but permitted option. For example, the use of the null encryption scheme for the SUCI, predictable allocation of temporary identifiers, or permissive modem handling of certain network messages.

Existing methodologies do not resolve this question. Traditional IMSI catchers rely on active jamming to force a target into rogue cells, a disruptive and readily detectable approach that does not provide insights about the vulnerability it exploits. Modem fuzzing injects malformed 5G New Radio (NR) signalling from a rogue base station with the goal of driving basebands into unstable or uncontrolled states. These techniques assume a Dolev-Yao adversary model, which implies that the attacker has full control over the RAN and Core networks during an extended period of time. Although valuable for discovering new vulnerabilities, targeting the implementation surface of the devices usually causes inherent service disruptions and remains visible for countermeasures, limiting their usage in real-world scenarios. In addition, conformance and differential-testing frameworks systematically discover where these modems deviate from the 3GPP state machine. However, they operate in controlled environments and are confined to static deployments.

To address these limitations, we present \newtool{}, a real-time 5G network security posture auditor that exploits the legitimate mobility procedure to evaluate security aspects of subscribers in the wild. \newtool{} proposes a methodology that manipulates the criterion of the cell-reselection process, a fundamental mobility procedure in cellular networks, to induce the subscriber to camp on a non-legitimate base station, establishing an attack opportunity to audit the current subscriber configuration with minimal service disruption and a low entry barrier. It solely builds from open-source stacks and inexpensive Software Defined Radio (SDR) hardware and implements a multi-vector methodology able to request subscriber identifiers, force Radio Access Technology (RAT) downgrade via crafted \textit{Registration Reject} cause codes, or induce blocking states into the subscriber's modem. For each vector, we attribute the root cause to design limitations or implementation gaps, experimentally showcasing that these attack surfaces affect the most advanced and secure iteration of mobile networks across different consumer devices through different operators.

Our contributions are the following:
\begin{itemize}
    \item \newtool{}, an open-source tool able to evaluate the posture of a subscriber operating on a real network. It pulls Commercial Off-The-Shelf (COTS) User Equipment (UE) subscribers into a self-created rogue cell from commercial 5G Standalone (SA) networks by exploiting the standardised cell-reselection procedure, performing the posture validation with minimal service disruption.
    \item A root-cause attribution of each enabled attack to protocol design or implementation gaps, clarifying which exposures survive full 3GPP compliance and which ones are operator/vendor-induced.
    \item An empirical characterisation of Global Unique Temporary Identifier (GUTI) allocation predictability across three commercial operators, quantifying the inter-registration step size and the resulting linkability of temporary identifiers that enables persistent tracking.
    \item The open-source release of the collected 5G datasets, including commercial signalling traces, rigorously anonymised and made available to the research community upon acceptance.
\end{itemize}

\section{BACKGROUND} \label{sec:background}

Despite the novel approaches included in the latest cellular generations, authentication material (e.g. subscriber identifiers) is key for the process, but is still exposed at various stages of the communication and malicious entities can still use them to succeed in executing multiple attack vectors~\cite{yomna2019gotta,cheng2023watching}. The following sections describe the role of the identifiers within the authentication process, their types and the management procedures required to update and protect them towards fulfilling the standard requirements~\cite{ts23003,ts23501,ts24501,ts33501}. Then, we describe the different IMSI catching techniques and their relation with the cell reselection and handover procedures.

\subsection{5G Identifiers}
\label{subsec:5gsec}

The International Mobile Subscriber Identity (IMSI) constitutes the cornerstone of subscriber identification within a cellular network. It is stored in the Universal Subscriber Identity Module (USIM) within the physical UE device and is externally known only by the Home Network (HN). Before the arrival of 5G, the IMSI was transmitted in plaintext during the initial steps of the registration procedure, making it susceptible to being intercepted by third-party observers (active or passive) in the wireless interface.

The inclusion of temporary identifiers (e.g., Temporary Mobile Subscriber Identity (TMSI) and Globally Unique Temporary Identifier (GUTI)) directly addressed this issue by limiting the interactions in which the IMSI was transmitted in clear text. Unfortunately, these mechanisms suffered from implementation flaws (e.g. low refresh rates), making them an alternative for device traceability~\cite{hong2018guti, Fact_Checking_5G_Security}. Furthermore, a Fake Base Station (F-BS) attacker can bypass temporary identifiers by requesting the IMSI directly from the user.

To effectively address these limitations, 5G was built upon two key concepts: i) avoiding transmitting the IMSI, replacing it with the Subscription Permanent Identifier (SUPI); and ii) adding encryption to the transmitted identifier, leading to an encrypted version of the SUPI called the Subscription Concealed Identifier (SUCI) to prevent exposing even the SUPI. The subscriber will generate the SUCI by using the HN cryptographic material (public key), which has been previously provisioned either directly in the USIM card or through secure procedures~\cite{ts23501,ts33501}. 

However, the effectiveness of the mechanisms above largely depends on the availability of legitimate 5G SA networks and their support for SUCI-based authentication, which, based on the results of recent works~\cite{Fact_Checking_5G_Security,ludant20235g,eleftherakis2024demystifying}, appears to be largely absent in current deployments that default to the ``Null-Scheme'' (cleartext) transmission.

Moreover, the 5G security model prioritises subscriber identity protection over equipment identity protection. This known design trade-off has privacy implications for subscribers, given that both the International Mobile Equipment Identity (IMEI) and its Software Version (IMEISV) can be used to authorise the hardware and are not protected by design. Furthermore, the fallback to previous generations' functionality improves the overall user experience but hinders the protections provided by 5G. The legacy architectures rely exclusively on symmetric cryptography and lack the public key infrastructure required to conceal the identifier before authentication. Consequently, any procedure that forces a device to downgrade or even reselect—to a 4G cell inherently strips away the SUCI protection, reverting the device to cleartext IMSI transmission.

Additionally, mobility procedures do not enforce the protection, which can drastically exacerbate the problem (e.g. device transitioning to a new cell through cell reselection or roaming). The following section describes the cell reselection procedures, emphasising the criterion followed by devices and identifying actions or states that can benefit an attacker to expose the subscriber identifier by impersonating a legitimate base station.

\subsection{5G Cell Reselection}
\label{subsec:5gcell-reselection}

The cell reselection process is a device-driven procedure that defines how a subscriber in \texttt{RRC\_IDLE} or \texttt{RRC\_INACTIVE} state selects the best cell to camp on (TS~38.304~\cite{ts38304}). Unlike the handover procedure, which is network-driven and requires the subscriber to be authenticated in \texttt{RRC\_CONNECTED} mode, cell reselection prioritises battery life and network scalability to the detriment of security. This architectural trade-off creates an inherent vulnerability, as the decision relies on unauthenticated broadcast parameters. Based on the assignment of priorities to operating frequencies~\cite{ts38304}, cell reselection is classified into three types:

\begin{itemize}
    \item[-] Inter-frequency: Cells that are on different frequencies within the same RAT. \textit{The attack vector described in this work exploits this procedure.}
    \item[-] Intra-frequency: Cells that are on the same frequency as the current serving cell (usually on the same RAT).
    \item[-] Inter-RAT: Cells using a different RAT (e.g. 5G SA to 4G LTE).

\end{itemize}

Thus, priorities can be received by the subscriber through System Information Blocks (SIB), dedicated signalling messages (e.g. RRCRelease) or inherited from another RAT (e.g. at inter-RAT selection). If the subscriber receives information from many of them, it must ignore all except the ones received through dedicated signalling. Then, measurement rules are also key during the reselection process to avoid overloading the subscriber with unnecessary measurements that will drain its battery. The subscriber will always collect information about frequencies with a higher priority than the current serving cell and, for those with equal/lower priority, when the current received signal is poor. The reselection process is performed based on the collected information and priorities about neighbouring cells, following the steps defined in Algorithm~\ref{alg:cell-reselection}. The core logic resides in holding three priority cases: 

\hfill

\noindent\textbf{[1] Higher Priority} The goal is to move the subscriber to a preferred frequency where the candidate cell exceeds a high-quality threshold (high frequency priority and quality signal strength).

\hfill

\noindent{\textbf{[2] Lower Priority}} The goal is to fall back to a lower priority frequency when the current service is failing. This occurs when the serving cell has low signal strength, and the candidate cell has a minimum signal quality specified by a threshold.

\hfill

\noindent{\textbf{[3] Equal Priority}} The main goal is to find the best cell among two candidates with equal frequency priority. This case avoids using thresholds to eliminate the possibility of ping-pong effects and computes the rank of the involved cells as follows:

\hfill

    \begin{itemize}
        \item Service cell rank: $R_{s} = Q_{meas,s} + Q_{hyst}$
        \item Candidate cell rank: $R_{c} = Q_{meas,c} - Q_{offset}$
    \end{itemize}
where $Q_{meas}$ is the signal strength (RSRP) or quality (RSRQ), $Q_{hyst}$ is a bonus added to the serving cell to make subscribers stay if possible, and $Q_{offset}$ is a penalty or bonus applied to the neighbour cell (often adjusted by the network to perform load balancing). Variables and parameters are summarized in Table~\ref{tab:reselection-variables}.

{\small
\begin{algorithm}[t!]
\caption{Cell Reselection Decision (TS 38.304)}
\begin{algorithmic}[1]
\Require Measurements of Serving Cell ($s$) and Candidate ($c$), Priorities
\Ensure Decide whether to reselect a new cell

\For{each $c$ in \textit{CandidateCells}}
    \State $T_{valid} \leftarrow (time > T_{\text{reselection}})$
\State \Comment{\textbf{Case\_1: Higher Priority (Absolute Priority)}}
    \If{$Prio_{c} > Prio_{s}$}
        \If{\textit{ThreshServingLowQ} is broadcast}
            \If{$Squal_{c} > Thresh_{c,HighQ} \land T_{valid}$}
                \State \textbf{Reselect} to $c$
            \EndIf
        \Else
            \If{$Srxlev_{c} > Thresh_{c,HighP} \land T_{valid}$}
                \State \textbf{Reselect} to $c$
            \EndIf
        \EndIf

\State \Comment{\textbf{Case 2: Lower Priority (Fallback)}}
    \ElsIf{$Prio_{c} < Prio_{s}$}
        \If{\textit{ThreshServingLowQ} is broadcast}
            \If{$Squal_{s} < Thresh_{s,LowQ} \land Squal_{c} > Thresh_{c,LowQ} \land T_{valid}$}
                \State \textbf{Reselect} to $c$
            \EndIf
        \Else
            \If{$Srxlev_{s} < Thresh_{s,LowP} \land Srxlev_{c} > Thresh_{c,LowP} \land T_{valid}$}
                \State \textbf{Reselect} to $c$
            \EndIf
        \EndIf

\State \Comment{\textbf{Case 3: Equal Priority (Ranking Rules)}}
    \ElsIf{$Prio_{c} == Prio_{s}$}
        \State Calculate Ranking $R_s$ and $R_c$ 
        \If{$R_c > R_s \land T_{valid}$}
             \State \textbf{Reselect} to $c$
        \EndIf
    \EndIf
\EndFor
\end{algorithmic}
\label{alg:cell-reselection}
\end{algorithm}
}

However, the described process still exposes certain flaws that can benefit an attacker. First, the cell reselection priorities are communicated through broadcast messages (SIBs) that can be manipulated by a non-legitimate base station towards directing the subscriber to reselect a specific and most probably untrusted cell. This limitation opens the possibility for IMSI catching and 
downgrade, given the power that the destination base station has, even without the need for the user to authenticate the network. Then, signal measurements-based decisions are prone to being manipulated by boosting signal power or the quality of an F-BS impersonating a candidate cell to satisfy the reselection conditions, leading to a similar situation. Finally, the lack of network authentication by the subscriber before the reselection enables a subscriber to camp on a cell which can be easily impersonated, and hence, enables tracking of users.

\begin{table}[h!]
\caption{Variables and Parameters for TS 38.304 Cell Reselection}
\label{tab:reselection-variables}
\centering
\footnotesize
\newcolumntype{Y}{>{\raggedright\arraybackslash}X} 
\begin{tabularx}{\linewidth}{@{} l Y @{}}
\toprule
\textbf{Variable} & \textbf{Description} \\ \midrule
$cell_{c}$, $cell_{s}$ & Candidate cell and Serving cell identifiers. \\
$Prio_{x}$ & Reselection priority assigned to freq. of cell $x$. \\
$\textit{ThreshServingLowQ}$ & \textbf{Control Parameter:} Forces RSRQ-based decision. \\
$Squal_{x}$ & \textbf{Signal Quality (RSRQ)} of cell $x$ (dB). \\
$Srxlev_{x}$ & \textbf{Signal Power (RSRP)} of cell $x$ (dBm). \\
$Thresh_{x,HighQ/P}$ & Threshold, \textbf{Higher Priority} cells (Q=Quality, P=Power). \\
$Thresh_{x,LowQ/P}$ & Threshold, \textbf{Lower Priority} cells (Q=Quality, P=Power). \\
$T_{\text{reselection}}$ & Time duration criteria must be met to trigger reselection. \\
$R_s, R_c$ & Ranking values (for Equal Priority) derived from RSRP. \\ \bottomrule
\end{tabularx}
\end{table}

Through the mechanisms highlighted above, an attacker can reliably manipulate the reselection process to force the UE to camp on an F-BS cell through an intra-cell or inter-cell procedure, enabling the collection of user identifiers or the downgrade of users to previous (more vulnerable) generations.

\begin{table*}[!ht]
  \centering
  \caption{Comparison of our work with the state-of-the-art regarding subscriber privacy (subscriber identifier capture), active traceability, and unauthenticated signalling resilience auditing in laboratory and commercial environments.}
  \label{tab:sota_comparison}
  
  \resizebox{\textwidth}{!}{%
    \begin{tabular}{l ccc ccc}
      \toprule
      
      \textbf{Reference} & 
      \multicolumn{3}{c}{\textbf{Laboratory 5G SA Environment}} & 
      \multicolumn{3}{c}{\textbf{Commercial 5G SA Environment}} \\
      
      \cmidrule(lr){2-4} \cmidrule(lr){5-7}
      
      & \textbf{Catching} & \textbf{Traceability} & \textbf{Unauth. Signalling} & \textbf{Catching} & \textbf{Traceability} & \textbf{Unauth. Signalling} \\
      
      \midrule

      \cite{ludant20235g}  
        & \cellcolor{lightred}\xmark 
        & \cellcolor{lightgreen}\cmark 
        & \cellcolor{lightred}\xmark 
        & \cellcolor{lightred}\xmark 
        & \cellcolor{lightgreen}\cmark 
        & \cellcolor{lightred}\xmark \\

      \cite{5gbasechecker}      
        & \cellcolor{lightgreen}\cmark 
        & \cellcolor{lightred}\xmark 
        & \cellcolor{lightgreen}\cmark 
        & \cellcolor{lightred}\xmark 
        & \cellcolor{lightred}\xmark 
        & \cellcolor{lightred}\xmark \\
      
      \cite{5Ghoul}      
        & \cellcolor{lightgreen}\cmark 
        & \cellcolor{lightred}\xmark 
        & \cellcolor{lightgreen}\cmark 
        & \cellcolor{lightred}\xmark 
        & \cellcolor{lightred}\xmark 
        & \cellcolor{lightred}\xmark \\
      
      \cite{still_catching_them_all} 
        & \cellcolor{lightgreen}\cmark 
        & \cellcolor{lightgreen}\cmark 
        & \cellcolor{lightred}\xmark 
        & \cellcolor{lightred}\xmark 
        & \cellcolor{lightred}\xmark 
        & \cellcolor{lightred}\xmark \\
        
      \cite{Never_Let_Me_Down}       
        & \cellcolor{lightred}\xmark 
        & \cellcolor{lightred}\xmark 
        & \cellcolor{lightgreen}\cmark 
        & \cellcolor{lightred}\xmark 
        & \cellcolor{lightred}\xmark 
        & \cellcolor{lightred}\xmark \\
        
      \cite{SNI5GECT}                
        & \cellcolor{lightgreen}\cmark 
        & \cellcolor{lightgreen}\cmark 
        & \cellcolor{lightred}\xmark 
        & \cellcolor{lightred}\xmark 
        & \cellcolor{lightred}\xmark 
        & \cellcolor{lightred}\xmark \\
      
      \midrule
      \textbf{\newtool{}} (Ours)     
        & \cellcolor{lightgreen}\cmark 
        & \cellcolor{lightgreen}\cmark 
        & \cellcolor{lightgreen}\cmark 
        & \cellcolor{lightgreen}\cmark 
        & \cellcolor{lightgreen}\cmark 
        & \cellcolor{lightgreen}\cmark \\
      
      \bottomrule
      
      \addlinespace[0.5ex]
      \multicolumn{7}{l}{\footnotesize \cmark: Features for collection of subscriber IDs, active traceability and behavioural information against unauthenticated} \\
      \multicolumn{7}{l}{ signalling messages covered.} \\
      \multicolumn{7}{l}{\footnotesize \xmark: Features not covered.} \\
    \end{tabular}
  }
\end{table*}

\subsection{Related Work}
\label{subsec:related_work}

Subscribers' identifier catching attacks (e.g. IMSI Catching) have evolved alongside cellular network technologies, intending to expose sensitive identifiers from network subscribers to perform location tracking, identification, communication interception or even communication denial. Starting from 4G LTE, Mjølsnes, Stig F., and Ruxandra F. Olimid (2017)~\cite{mjolsnes2017easy} successfully showcased the feasibility and simplicity of IMSI catching in 4G networks through unmodified open-source solutions and generic off-the-shelf hardware. The authors not only successfully collected subscriber identifiers but also made the network unavailable to the device (denial of service). Fortunately, IMSI catchers have been demonstrated to be useful in disaster scenarios such as the one described by Antonio Albanese et al. (2021)~\cite{sardo}, which enables identity-awareness use cases such as locating targets for public safety.

The arrival of 5G brought different works starting from Chlosta, Merlin et al. (2021)~\cite{still_catching_them_all}, whose authors demonstrated the feasibility of performing SUCI catching in commercial 5G Standalone (SA) networks. However, their experimental results on commercial networks are limited to 4G networks, even though they claim that the obtained results shall be extensible to 5G. Ludant, Norbert, et al. (2023)~\cite{ludant20235g} propose the first real-time control channel sniffer for 5G-NR capable of inferring encrypted RRC information (e.g. scrambling ID of users from information leaks). The results show that users can be tracked within a certain area using the Radio Network Temporary Identifiers (RNTIs), thanks to the decoding capabilities provided by the control channel sniffer. However, the proposed attack requires an active interaction through specific applications (e.g., Signal or Telegram) to generate predictable patterns that the control channel sniffer can recognise. The retrieved information from the subscriber is rather limited. 

Karakoc, Bedran, et al. (2023)~\cite{Never_Let_Me_Down} focus on downgrade attacks whose purpose is to make the target device move to a previous, less secure, network generation (e.g. from 5G to 3G). In addition to the proposed systematic testing framework, the authors propose mainly two case studies: i) the ``5G SA to 2G downgrade'', where they demonstrate that even a phone connected to a highly secure 5G SA network can be downgraded to 2G; and the ``5G NSA Null-Encryption Bid-Down'', where they push a no encryption supported message to the network resulting in an unencrypted 5G data session. However, the results in commercial networks are limited to showcasing that the Anti-bidding-down parameter is not currently being implemented by real deployments, while the most ``aggressive'' attacks, such as the previously mentioned test cases, were only tested in a laboratory environment. Luo, Shijie, et al. (2025)~\cite{SNI5GECT} proposed the first framework that can sniff 5G traffic in real-time and inject malicious packets into legitimate connections without the need for F-BSs by tracking the exact state of the connection (e.g. to inject a malformed packet or a \textit{Registration Reject}). However, as distance increases ($>$20~m), the accuracy of the sniffing and injection drastically drops, which, in combination with environmental interferences, could drastically decrease the feasibility of the attack in real-world scenarios.

A complementary line of work automates the discovery of baseband implementation flaws. Garbelini, Matheus E., et al. (2025)~\cite{5Ghoul} present \textit{5Ghoul}, a stateful multi-layer fuzzing framework that injects malformed RRC and NAS messages over the air from a F-BS, driving commercial 5G modems into unstable states (e.g. crashes, hangs and involuntary downgrades) that typically require a manual reboot. Tu, Kai, et al. (2024)~\cite{5gbasechecker} propose \textit{5GBaseChecker}, a black-box framework that infers finite-state-machine models of baseband control-plane behaviour and applies differential testing across devices to expose deviations from the 3GPP specification, uncovering numerous exploitable non-compliances across basebands from multiple vendors. While both works are highly effective at finding implementation vulnerabilities, they share three properties that separate them from an in-the-wild security auditing: i) they operate on devices under test within controlled testbeds rather than against devices living in commercial deployments; ii) they rely on malformed or fuzzed inputs that are inherently service-disruptive and detectable; and, iii) by construction, they target the implementation surface, without considering (potentially harming) behaviours that may be caused by a deployment/vendor choice. As depicted in Table~\ref{tab:sota_comparison}, both tools have the features to perform catching and unauthenticated signalling auditing, but none of them provides results on devices operating commercial 5G SA networks.

Along with the works focusing on exploiting vulnerabilities of current cellular networks, the advent of 5G also brought different works that focus on checking the current security status of commercial networks, disclosing possible gaps or misconfigurations that may harm the overall system security. Lasierra, Oscar, et al. (2023,2025)~\cite{European_5G_Security_in_the_Wild, Fact_Checking_5G_Security} and Eleftherakis, Stavros, et al. (2024)~\cite{eleftherakis2024demystifying} provided an extensive study of commercial networks in terms of security, reporting a poor status of commercial networks implementing the actual security features proposed by 5G standards. They also highlight a possible lack of randomness in temporary identifiers, which exacerbates the problem of user traceability. Their results showcase the limitations in current implementations, given that many critical security features are marked as optional in the standards.

Finally, the majority of the abovementioned works use F-BS to fully or partially perform certain attacks, and hence, different works have been trying to disclose the vulnerabilities of fundamental radio-level protocols (e.g. cell selection and broadcast messages) and usually evaluate tools for F-BS identification. Park, Shinjo. (2023)~\cite{park2023we} provides a comprehensive study on how advanced features introduced by 5G (e.g. SUCI) affect IMSI-catching and evaluates existing detection tools. The results show that novel security features are not directly affecting IMSI catchers because they are essentially based on detection, and as detection gets better, attackers use smarter tactics such as low-power cloning or partial protocol execution to avoid standard detection mechanisms. Paci, Andrea, et al. (2025)~\cite{FlashCatch} revealed that the weakest link of IMSI catchers is not the identification but rather the disruption it causes to the devices connecting to them (e.g. several seconds of connection loss). To avoid it, the authors collect IMSI/TMSI in sub-second time by sending an \textit{Authentication Failure} to the target device, making it think that the cell is having issues and immediately return to a legitimate network. 

While prior research either highlights fragmented vulnerabilities restricted to laboratory environments, relies on active user traffic, automates implementation-flaw discovery on devices under test through disruptive fuzzing or differential testing, or limits itself to passive network auditing as summarised in Table~\ref{tab:sota_comparison}, there remains a critical need for a comprehensive, low-disruption evaluation of F-BS-based attack viability in real-world 5G deployments that explicitly attributes each exposure to an inherent protocol design limitation or an operator/vendor implementation gap.

\section{METHODOLOGY}
\label{sec:methodology}

\subsection{Threat Model}
\label{subsec:threat_model}

The primary objective of the adversary is to harvest sensitive information (e.g. permanent/temporary identifiers) of victims within a specific coverage area. This is achieved by operating an F-BS capable of impersonating a legitimate commercial cell, manipulating the cell reselection criteria to force nearby subscribers to camp on the malicious node. To successfully execute this attack, the F-BS must: (1) Broadcast parameters consistent with the target commercial network. Specifically, the Mobile Country Code (MCC) and Mobile Network Code (MNC) must match those of the legitimate operator. (2) Acquire knowledge of the commercial network's frequency plan. To this end, the attacker analyses the \textit{SIB4} broadcast by the legitimate network. This reveals the cell reselection priorities assigned to neighbouring cells. (3) Satisfy the signal quality thresholds required for reselection. While the \textit{Absolute Priority} mechanism minimises the need for high transmission power, the F-BS signal must still exceed $Thresh_{High}$ at the victim's location. 

\begin{figure*}[!ht] 
    \centering
    \includegraphics[width=0.95\textwidth, alt={Flow diagram illustrating a F-BS attack}]{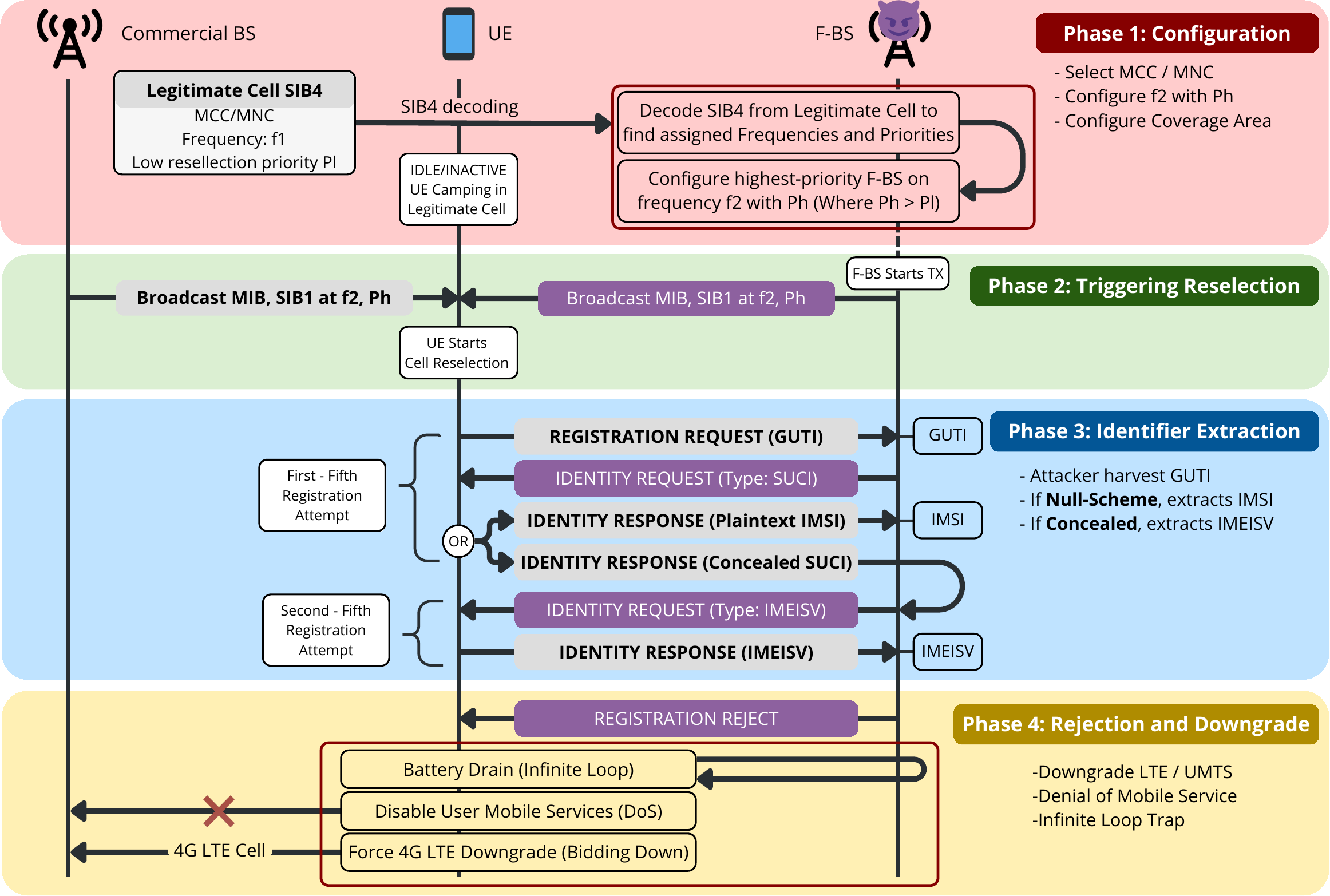}
    \caption{Flow diagram of a fake base station attack vector implementation}
    \label{fig:Flow_diagram_fake_BS_attack}
\end{figure*}

\subsubsection*{\textbf{Target Definition}}
The target is defined as any subscriber of the commercial 5G network located within the effective coverage of the F-BS. We assume the victim uses a standard COTS UE with no hardware or software modifications. The UE strictly follows the 3GPP TS 38.304~\cite{ts38304} mobility standard.

Crucially, the attack is transparent to the user, meaning that no interaction is required and no security warnings are displayed. The device autonomously performs a standard reselection to the F-BS and initiates the \texttt{REGISTRATION} procedure, a situation that may result in IMSI exposure in the subsequent \texttt{IDENTITY\_RESPONSE} message. The disruption to legitimate service is designed to be minimal and transient (see Section~\ref{sec:evaluation}) so as not to require user interaction or raise any OS-level security warning, though transient connectivity loss may occur.

\subsubsection*{\textbf{Operational Constraints}}
The attack is strictly feasible only when the target device is in \texttt{RRC\_IDLE} or \texttt{RRC\_INACTIVE} mode as described in Section~\ref{subsec:5gcell-reselection}. In these states, the radio connection is released to conserve power, and mobility is device-controlled (Cell Reselection). The subscriber periodically wakes up to monitor Master Information Blocks (MIBs), System Information Blocks (SIBs), and Paging messages, creating the window of opportunity for the F-BS to influence the cell selection decision. In order to infer the available time window for the F-BS to operate, we measure the time a regular device was in \textit{RRC\_IDLE} state using ``adb shell dumpsys'' in our devices, showing that almost 70\% of the time, the device remained in that state.

\subsection{Attack Vector Overview}
\label{subsec:attack_overview}

The proposed attack vector exploits three protocol-design limitations: i) 
subscriber cell-reselection procedure, ii) unauthenticated \texttt{IDENTITY\_REQUEST}
acceptance, and iii) unauthenticated \texttt{REGISTRATION\_REJECT} message processing to audit the security posture followed by the subscriber while operating a real network.

Besides the cell reselection threat model described in Section~\ref{subsec:threat_model}, the proposed attack vector exploits the network's ability to request the permanent identifier (SUCI/IMSI) via an \texttt{IDENTITY\_REQUEST} message before any mutual authentication or key exchange has occurred. Furthermore, the standard permits the network to issue unauthenticated \texttt{REGISTRATION\_REJECT} messages with various cause codes, allowing the attacker to manipulate the device's future states or force a downgrade to legacy generations (4G/3G). Additionally, the user provides the attacker with the previously assigned temporary identifier (GUTI) in the initial \texttt{REGISTRATION\_REQUEST} message after cell reselection. This deterministic, four-phase attack execution is detailed below, with the corresponding message flow depicted on the right side of Fig.~\ref{fig:Flow_diagram_fake_BS_attack}.

\subsubsection*{\textbf{Phase 1:} Configuration}
\label{subsubsec:Phase_1}
The process begins with the attacker choosing the commercial 5G SA vendor (MNC) within the respective country (MCC). Then, by decoding \textit{SIB4} messages from the legitimate serving cell, the adversary identifies the frequency bands with the assigned reselection priority. After that, the F-BS is configured to impersonate a commercial cell on the highest-priority frequency (left side of Figure~\ref{fig:Flow_diagram_fake_BS_attack}). This configuration ensures that the malicious cell is not merely a valid candidate but a \textit{preferred} one according to the standard reselection logic (See Section~\ref{subsec:5gcell-reselection}).

\subsubsection*{\textbf{Phase 2:} Triggering Reselection}
Once the F-BS is active, the attack exploits the frequency priority logic. Unlike traditional jamming~\cite{mjolsnes2017easy} or overshadowing attacks~\cite{SNI5GECT} that require the F-BS to physically overpower the legitimate serving cell, our approach relies on the subscriber's compliance with 3GPP TS 38.304. If the F-BS broadcasts a higher priority (or equal priority with sufficient signal quality), the subscriber is mandated to reselect it as long as the signal exceeds the minimum threshold ($Thresh_{High}$). This allows the attack to succeed with lower transmission power, potentially evading detection systems based on signal strength anomalies~\cite{fbsli2017fbs,fbshuang2018identifying}.

Upon activation, three outcomes are possible for a target subscriber in the vicinity: (1) \textbf{Failure (Different 5G Cell)}, the subscriber camps on a different legitimate 5G cell (e.g., if the F-BS signal is below $Thresh_{High}$). (2) \textbf{Partial Failure (Inter-RAT Fallback)}, the subscriber camps on a legitimate 4G cell due to signal fluctuations. This effect can be mitigated by adjusting the configured priority in the F-BS. (3) \textbf{Success (F-BS Camping)}, the subscriber selects the F-BS and initiates the registration procedure.

\subsubsection*{\textbf{Phase 3:} Identifier Extraction}
Once the subscriber camps on the F-BS, it transmits a \texttt{REGISTRATION\_REQUEST} of type \textit{Mobility Updating}, containing its temporary identity (5G-GUTI). The F-BS intercepts this and immediately responds with an \texttt{IDENTITY\_REQUEST} (Type: SUCI), claiming that the provided GUTI is unrecognised (subscriber context no longer valid). Consequently, the target responds with its SUCI. If the target posture of the network utilises the \textit{Null-Scheme} (Protection Scheme 0), this SUCI contains the IMSI in plaintext. Conversely, if the selected posture is either \textit{Profile A} or \textit{Profile B}, the subscriber correctly transmits a cryptographically concealed SUCI. During operation, the attacker can seamlessly pivot by issuing a subsequent \texttt{IDENTITY\_REQUEST} explicitly demanding the device's IMEI/IMEISV.

\subsubsection*{\textbf{Phase 4:} Rejection and Downgrade}
Upon successful extraction of the target identifiers, the F-BS terminates the connection by transmitting an unprotected \texttt{REGISTRATION\_REJECT} message. By strategically selecting specific 5G Mobility Management (5GMM) cause codes (defined in TS 24.501~\cite{ts24501}), the attacker can exploit the device's error-handling state machine to dictate its subsequent behaviour. Some of the outcomes may instruct the subscriber to: delete its current 5G security context and initiate a fresh registration procedure, force the device to abandon the 5G network entirely and fall back to legacy LTE architectures (Bidding-Down), or strictly prohibit the subscriber from accessing connectivity services, triggering a persistent DoS.

\section{\newtool{} IMPLEMENTATION SETUP} \label{sec:impl}

\newtool{} prototype was developed to audit the security of subscribers and categorise the vulnerabilities while operating both 5G SA and Non-Standalone (NSA) mobile networks. The architecture utilises a unified hardware infrastructure capable of switching between 5G SA and NSA technologies using separate software stacks.

\subsection{\textbf{Hardware and Software}}
\label{subsec:impl-sw_hw}
\begin{figure}[h] 
    \centering
    \includegraphics[width=0.45\textwidth, alt={Setup image}]{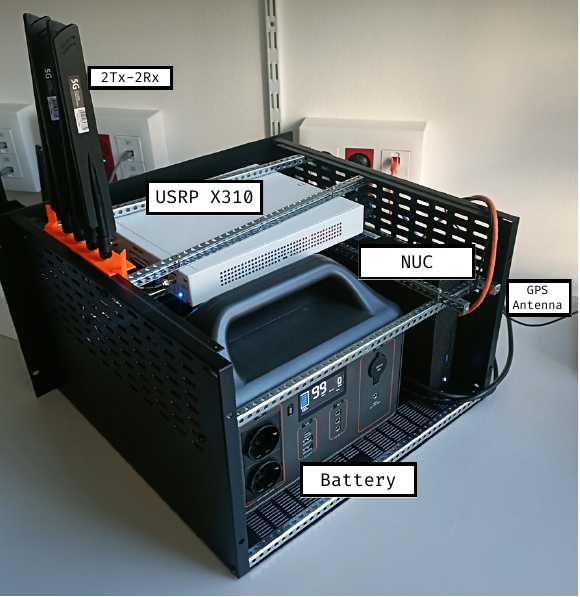}
    \caption{\newtool{} physical composition}
    \label{fig:Setup}
    \vspace{-4mm}
\end{figure}

The core \newtool{} logic is built upon the srsRAN 5G~\cite{srsran5g} (gNodeB) and srsRAN 4G~\cite{srsran4g} (eNodeB and EPC) projects. For the 5G SA Core Network (CN), we utilised Open5GS~\cite{open5gs} and introduced custom modifications to the Access and Mobility Management Function (AMF) to bypass authentication and manipulate signalling messages. All of it has been integrated into a module (Fig.~\ref{fig:Setup}) comprising a USRP X310 SDR, selected for its high-bandwidth capabilities and superior signal quality. For the processing unit, an Intel NUC equipped with a 13th~Gen Intel Core i9-13900H processor and 32~GB of RAM has been chosen. The module has 4 antennas (2 receiving and 2 transmitting channels) and one GPS antenna connected to the X310 SDR. High-bandwidth communication between the NUC and the USRP X310 is established via a direct 10~GbE interface, ensuring stable sample streaming.

For the target UE, we have used seven COTS devices with 5G SA technology support listed in Table~\ref{tab:imsi_attack}. To establish a controlled baseline for evaluating the selected subscribers, we deployed a private srsRAN 5G SA network and provisioned each device with a custom, configurable SIM card so that they can access the network as real customers.

Finally, to retrieve the network parameters required to configure \newtool{} in Phase 1 (\ref{subsubsec:Phase_1}), we passively extracted Absolute Radio Frequency Channel Numbers (ARFCNs) and SIB4 data using various COTS diagnostic tools, ranging from deep packet inspection software (Keysight Nemo Handy) \footnote{\url{https://www.keysight.com/us/en/product/NTH50047B/nemo-handy-handheld-measurement-solution.html}} to native OS engineering interfaces (Samsung Service Mode: $*\#0011\#$ dial).

\subsection{Attack Vector Execution}
\label{subsec:impl_attack_vectors}

\newtool{} handles the RF transmission and standard priority-based reselection mechanisms (Phases 1 and 2 described in Sec.~\ref{subsec:attack_overview}) using the srsRAN software suite (gNB). Moreover, the core logic of the attack vector is performed through a modified version of \textit{Open5GS CN}. 

Specifically, the AMF was reprogrammed to automate the identifier extraction and forced downgrade mechanisms (Phases 3 and 4 described in Sec.~\ref{subsec:attack_overview}). Upon receiving a \texttt{REGISTRATION\_REQUEST} from a lured subscriber, the modified AMF bypasses standard context retrieval. It intentionally fails the 5G-GUTI resolution process and immediately generates an \texttt{IDENTITY\_REQUEST} demanding the SUCI/IMSI.

Following the receipt of the \texttt{IDENTITY\_RESPONSE}, the rogue AMF logs the identifier, and it can be dynamically reprogrammed to subsequently demand the device's IMEI or IMEISV instead of the SUCI. To terminate the attack cycle as fast as possible to minimise the disruption, \newtool{} responds with a \texttt{REGISTRATION\_REJECT}. The AMF's rejection logic was designed to dynamically inject specific \textit{5GMM} cause codes into this unauthenticated message. By toggling these parameters, \newtool{} can seamlessly switch between executing an ``Implicitly De-registered'' loop (Cause \#10) to facilitate continuous identifier harvesting and bidding-down attacks or a ``5GS Services Not Allowed'' ban (Cause \#7) to trigger an immediate Denial of Service (DoS).

The following steps summarise the execution of the attack vector implemented by \newtool{} for auditing the posture of a subscriber:
\begin{itemize}
\item \textbf{Phase 1 --- Configuration:} The \textit{srsRAN} gNodeB is configured with parameters mimicking three different operators across five distinct configurations (three NSA and two SA network profiles).
\item \textbf{Phase 2 --- Triggering Reselection:} Utilising standard priority-based reselection triggers, the system employs Time-Division Multiplexing (TDM) to rotate between operators in one-minute windows. This ensures that the attack window is sufficient to trigger reselection for all users associated with the targeted networks.
\item \textbf{Phase 3 --- Identity Extraction:} Within the active TDM window, the modified \textit{Open5GS} AMF bypasses standard context retrieval. By forcing a failed GUTI resolution, the AMF compels the subscriber to disclose their SUCI/IMSI or IMEISV.
\item \textbf{Phase 4 --- Rejection/Downgrade:} The parameterised AMF logic injects the selected \textit{5GMM} cause codes, determining whether the subscriber remains in a retry loop or is barred from 5G services.
\end{itemize}

In practice, \newtool{} was configured to systematically impersonate three Tier-1 mobile network operators (designated herein as Operators A, B, and C). While all three providers maintained 5G NSA coverage in the testing area, 5G SA configurations were limited to Operators A and B, accurately reflecting the local deployment status of commercial infrastructure at the time of testing. \newtool{} broadcasts the specific network parameters (Public Land Mobile Network (PLMN) ID, frequencies, and priorities) of a single target operator for a custom, configurable duration. Empirical testing demonstrated that a one-minute window ($58~s$ on average) is sufficient to trigger reselection, gather idle-mode subscribers subscribed to that specific network, and complete the registration loops so that the subscriber can reconnect to the commercial legitimate network. The system can seamlessly rotate to another operator configuration, resulting in a continuous, automated cyclic loop.

During each cycle, \newtool{} extracts a comprehensive set of telemetry and identity data to perform the posture audit. Beyond the critical permanent (IMSI/SUCI) and temporary (GUTI) identifiers, the system logs the subscriber's supported security algorithms, precise network disconnection timestamps, the target's Home Public Land Mobile Network (HPLMN: Country and Provider), and the frequency of \newtool{} registration attempts, which enables localized device tracking within the targeted coverage area.

\section{EXPERIMENTAL EVALUATION} 
\label{sec:evaluation}

In this section, we present the experimental evaluation and results of \newtool{}. First, we validate the behaviour of \newtool{} and verify the 5G SA capabilities of each device under test. Second, we perform two different campaigns devoted to auditing subscriber privacy in both 5G SA and NSA scenarios by using \newtool{} to gather permanent and temporary identifiers, concluding with a study about subscriber traceability. Lastly, we conduct a campaign with \newtool{} implementing the full attack vector to audit the subscriber behaviours to unexpected \textit{Registration Reject} codes (including downgrade). The last two scenarios were performed in real-world scenarios targeting subscribers owned by our research team but fully connected to commercial 5G networks.

\subsection{Laboratory Setup: 5G SA}
\label{subsec:setup}

In the first scenario, we deployed in our laboratory a complete 5G SA Network using the open-source srsRAN project for gNB in conjunction with Open5GS for the 5G CN, allowing us to fully control the network behaviour and use COTS UEs provisioned with programmable SIM cards \footnote{\url{https://sysmocom.de/products/sim/sysmoisim-sja5/index.html}}. Within this controlled environment, we introduced \newtool{}, simulating adversarial conditions while ensuring the repeatability of the experiments that will be performed in the wild. This setup allowed us to validate the actual SA capabilities of various terminals assembling SA modems from different manufacturers implementing different 3GPP releases. 

Table~\ref{tab:imsi_attack} summarises the evaluation results in the ``srsRAN SA Check'' column. As depicted, six out of seven devices connected to our laboratory 5G SA network, confirming the devices' SA capabilities. However, \textit{OnePlus 8} is not able to successfully connect to the SA network. We further investigate this behaviour by using the Android Debug Bridge (adb) tool to check the current configuration of the device and \textit{vendors.xml}, a configuration or data file written in XML format that holds specific settings, details, or rules related to a hardware or software supplier. The results show that the 5G option is never reported at the framework level (e.g. through the \textit{Service State} events observed) and is not available at the user interface (within the preferred networks selection list). This configuration is not stored within the general rules to be applied for different operators within the \textit{vendors.xml} file, leaving the root cause to the OxygenOS framework. For this reason, we exclude this device from the subsequent audits.

\subsection{Subscriber Privacy Audit}
\label{subsec:subscriber-privacy-audit}

\subsubsection{\textbf{Privacy audit in commercial NSA scenarios}}
\label{subsec:wild_nsa}

To evaluate the feasibility of identity exposure in a realistic setting, we transitioned our experiments to a real-world setting by deploying \newtool{} configured with commercial 5G network parametrisation covering the laboratory space. 

\hfill

\noindent{\textbf{Configuration:}}
As previously described, \newtool{} combines srsRAN software, an enhanced version of Open5GS and an SDR device. The rogue entity was configured to broadcast a System Information Block Type 1 (SIB1) with a distinct Tracking Area Code (TAC) not present in the subscriber's current Tracking Area Identity (TAI) list, alongside a higher RSRP relative to the legitimate commercial network to trigger, as described in Section~\ref{subsec:attack_overview}, an immediate cell reselection.

\begin{algorithm}[h]
\caption{5G NSA IMSI Catching}
\label{alg:lte_attack}
\begin{algorithmic}[1]
\State \textbf{Attacker:} Broadcasts SIB1 with Highest Priority, $TAC_{fbs}$ and sufficient $RSRP$
\State \textbf{UE:} Performs Cell Reselection $\rightarrow$ \newtool{} Cell
\State \textbf{UE:} Sends \texttt{NAS\_TAU\_REQUEST(GUTI)}
\State \textbf{Attacker:} Sends \texttt{NAS\_TAU\_REJECT(Cause=\#10)}
\State \textbf{UE:} Deletes Security Context ($GUTI, K_{ASME}$)
\State \textbf{UE:} Sends \texttt{NAS\_ATTACH\_REQUEST}
\State \textbf{Attacker:} Sends \texttt{NAS\_IDENTITY\_REQUEST(IMSI)}
\State \textbf{UE:} Sends \texttt{NAS\_IDENTITY\_RESPONSE(IMSI)}
\State \textbf{Attacker:} \textit{Log IMSI and Security Caps; Drop Connection}
\State \textbf{UE:} $T3410$ Timeout $\rightarrow$ Bar \newtool{} Cell $\rightarrow$ Fallback Legit. Net.
\State \textbf{UE:} $T3402$ Timeout $\rightarrow$ Unbar \newtool{} Cell $\rightarrow$ \textbf{Goto Step 1}
\end{algorithmic}
\end{algorithm}

\noindent{\textbf{Attack Execution Flow:}}
The sequence observed is depicted in Algorithm~\ref{alg:lte_attack} and proceeded as follows:

\begin{enumerate}
    \item[(1)] \textbf{Initial Attraction:} The target subscriber, detecting prioritised frequency jointly with the necessary signal quality of \newtool{} configured with a  new TAC, initiates a \texttt{Tracking Area Update (TAU) Request}. This message is integrity-protected using the subscriber's security context created in the legitimate network.
    
    \item[(2)] \textbf{Context Invalidation (The Exploit):} Since \newtool{} does not possess the valid cryptographic keys ($K_{ASME}$) to decrypt or verify the message, it actively rejects the request. Crucially, the rejection is sent with \textbf{EPS Mobility Management (EMM) Cause Code \#10 (Implicitly Detached /Implicitly De-Registered)}, which, as described in Section~\ref{subsec:attack_overview}, forces the subscriber to re-attempt registration.
    
    \item[(3)] \textbf{Forced Re-Attachment:} Upon receiving Cause \#10, the standard mandates the subscriber to delete its current GUTI and security context. The subscriber immediately attempts to re-register by sending an \texttt{Attach Request}.
    
    \item[(4)] \textbf{Identity Extraction:} The subscriber identifies itself using its last known GUTI. \newtool{} responds with an \texttt{Identity Request} (type: IMSI). Due to the lack of pre-authentication in LTE (control plane for 5G NSA), the subscriber complies and transmits an \texttt{Identity Response} containing the permanent \textbf{IMSI} in cleartext.
    
    \item[(5)] \textbf{Intelligence Gathering:} In addition to the IMSI, the \texttt{Attach Request} message exposes the subscriber's \textbf{Network Capabilities}, revealing supported integrity and encryption algorithms (e.g., 128-EEA1, 128-EEA2, or lack thereof), allowing device profiling for potential downgrade attacks or to create a virtual fingerprint of the device to increase precision on user tracking attacks.
\end{enumerate}

\hfill

\noindent{\textbf{Results:}} Our results show that all devices exposed their permanent identifiers for all the studied operators during the execution of the attack vector. These results are represented as white circles (IMSI exposed) for each combination of device model and operator in Table~\ref{tab:imsi_attack} within the \textit{NSA IMSI} column. Given that NSA networks still rely on the 4G control plane, the vulnerability highlighted here is a \texttt{protocol-design limitation}.

\noindent\textbf{Timer-Induced Anomalies: }Following the successful exfiltration of the IMSI, we decided to further explore the behaviour of subscribers when connected to \newtool{} and hence, it was configured to withhold further signalling (i.e., sending neither \textit{Attach Accept} nor \textit{Reject}). The observed behaviour highlights a silence timer expiration (T3410) with the following consequences:

\begin{itemize}
    \item \textbf{Temporary Barring:} The subscriber marks the \newtool{} cell as temporarily barred and reverts to the legitimate commercial network.
    
    \item \textbf{The Loop:} We observed that after an implementation-dependent back-off period (governed by timer T3402, typically 12 minutes, though often shorter in specific firmware implementations), the subscriber unbars the \newtool{} cell. Since the \newtool{} signal remains dominant, the subscriber re-selects it, and the extraction cycle repeats, allowing for continuous tracking of the target device.
    
\end{itemize}

\subsubsection{\textbf{Privacy audit in commercial SA scenarios}}
\label{subsec:wild_sa}

This evaluation scenario targets 5G SA environments. Unlike the 4G scenarios, 5G SA introduces stronger privacy preservation mechanisms (e.g. SUCI). However, we demonstrate that active manipulation of radio conditions and frequency priorities within our controlled lab environment can still force the subscriber into an insecure state.

\noindent{\textbf{Configuration:}}
The target subscriber was initially camping on a legitimate commercial 5G SA network for Operator A, which assigned a cell reselection priority of 5 (where 7 is the maximum). In contrast, Operator B configured all frequencies within the controlled area with the maximum priority (7). This configuration data was collected by analysing SIB4 messages for all operators under study. Unfortunately, Operator C does not provide 5G SA coverage within the test area. Consequently, SA IMSI catching was excluded from this specific evaluation.

The configuration employed by Operator A represents an ideal scenario for an attacker, as the target cell's lower priority allows the malicious station to be configured with a maximum priority (7), triggering an immediate cell reselection based on 3GPP priority criteria. In contrast, Operator B presents a more complex challenge. By assigning the maximum priority to all intra-frequency cells, the subscriber is not forced to re-select based on priority alone (see Section~\ref{subsec:5gcell-reselection}). To overcome this equal-priority constraint without increasing transmission power, we manually degraded the legitimate signal quality. This was achieved by applying electromagnetic shielding (e.g., aluminium foil) to attenuate the commercial signal while maintaining a line-of-sight (LoS) path to the controlled environment where the experimental tool was operating.

\begin{algorithm}[h]
\caption{5G SA IMSI Catching}
\label{alg:5g_attack}
\begin{algorithmic}[1]
\State \textbf{Initial State:} UE connected to Legitimate SA Cell with $f_1$ and High Reselection priority ($P_1$)
\State \textbf{Action:} Move UE to poor coverage 
\State \textbf{Attacker:} Active F-BS on $f_2$ with High Reselection priority ($P_2$)
\State \textbf{UE:} Detects F-BS $f_2$ signal $> S_{nonIntraSearch}$
\State \textbf{UE:} Performs Cell Reselection $\rightarrow$ F-BS (Due to $R_c > R_s \land T_{valid}$) $P_1$ = $P_2$ (Case 3, Equal Priority, algorithm~\ref{alg:cell-reselection})
\State \textbf{UE:} Sends \texttt{NAS\_REGISTRATION\_REQUEST(GUTI)}
\State \textbf{Attacker:} Sends \texttt{NAS\_IDENTITY\_REQUEST(IMSI)}
\State \textbf{UE:} Sends \texttt{NAS\_IDENTITY\_RESPONSE(IMSI Cleartext)}
\State \textbf{Attacker:} Sends \texttt{NAS\_REGISTRATION\_REJECT}
\State \textbf{Burst Loop:} Steps 6---9 repeat \textbf{5 times} (Max Retries Reached)
\State \textbf{UE:} Bars F-BS Cell $\rightarrow$ Starts \textbf{T3502} (12 min) $\rightarrow$ Fallback to Legitimate 4G/NSA
\State \textbf{UE:} During T3502 wait in 5G NSA legitimate network as fallback $\rightarrow$ Reselects/Camps on F-BS \textbf{$f_2$ ($P_2$)}
$\rightarrow$ \textbf{Goto Step 6}
\end{algorithmic}
\end{algorithm}

\noindent{\textbf{Attack Execution Flow:}}
The sequence of steps to perform this attack is summarised in Algorithm~\ref{alg:5g_attack}. The reselection procedure of the subscriber is manipulated by influencing the reselection criteria and frequency hierarchy of the process. These actions make the subscriber select the cell broadcast by \newtool{} actively: 

\begin{enumerate}
    \item[(1)] \textbf{Initial Attraction:} We apply physical attenuation at the place where \newtool{} coverage was available within the controlled area.
    \begin{enumerate}
        \item[1.1] Following the standard mobility procedures, the subscriber performed a reselection to the legitimate cell, which offers better coverage. Once the subscriber was stable on this cell, we removed the physical attenuation so that the subscriber is now within the controlled area.
        \item[1.2] Detecting the strong signal from \newtool{}, the subscriber immediately initiated a cell reselection procedure to attach to what it perceived as the ``best'' available cell.
        \item[1.3] The subscriber is now camping on the cell advertised by \newtool{}.
    \end{enumerate}
    
    \item[(2)] \textbf{Context Invalidation (The Exploit):} The subscriber transmitted a \texttt{Registration Request} including its 5G-GUTI, which will be collected by \newtool{} but disregarded for the registration procedure because our objective is permanent identifiers.
    
    \item[(3)] \textbf{Forced Re-Attachment:} \newtool{} subsequently sends a \texttt{NAS Identity Request} explicitly querying for the permanent identity (SUPI/IMSI) type for correct authentication.
    
    \item[(4)] \textbf{Identity Extraction:} The subscriber complies with the standard and replies with the requested identity, which can be protected if the correct security profile is active or in clear text if the ``NULL Scheme'' profile is active.
    
    \item[(5)] \textbf{Intelligence gathering:} Similarly to the attack on NSA deployments (Section~\ref{subsec:wild_nsa}), temporary identifiers and subscribers' security capabilities are captured by \newtool{}, enabling device fingerprinting and tracking attacks.
\end{enumerate}

\begin{table*}[!ht]
  \centering
  \caption{Feasibility of the IMSI catching attack for different ME (left) under different SIM Card Providers (top).}
  \label{tab:imsi_attack}
  
  \resizebox{\textwidth}{!}{%
    \begin{tabular}{llc c cc cc cc}
      \toprule
      
      \textbf{Mobile Equipment} & 
      \textbf{Modem} & 
      \textbf{Rel.} & 
      \textbf{srsRAN} & 
      \multicolumn{2}{c}{\textbf{Operator A}} &
      \multicolumn{2}{c}{\textbf{Operator B}} &
      \multicolumn{2}{c}{\textbf{Operator C}} \\
      
      \cmidrule(lr){3-3} \cmidrule(lr){4-4} \cmidrule(lr){5-6} \cmidrule(lr){7-8} \cmidrule(lr){9-10}
      
      (ME) Model& & (3GPP) & 
      \textbf{SA} & 
      \textbf{SA} & \textbf{NSA} &
      \textbf{SA} & \textbf{NSA} &
      \textbf{SA} & \textbf{NSA} \\
      
      & & & 
      \textbf{Check} & 
      \textbf{IMSI} & \textbf{IMSI} & 
      \textbf{IMSI} & \textbf{IMSI} &
      \textbf{IMSI} & \textbf{IMSI} \\
      \cmidrule(lr){1-1} \cmidrule(lr){2-2} \cmidrule(lr){3-3} \cmidrule(lr){4-4} \cmidrule(lr){5-6} \cmidrule(lr){7-8} \cmidrule(lr){9-10}

      Galaxy Z Flip3 & Snapdragon X60  & 15  & \cmark & \xmark        & \ding{109} & \xmark      & \ding{109} & \xmark & \ding{109} \\
      OnePlus8        & Snapdragon X55  & 15  & \xmark & - & - & - & - & - & - \\
      Oppo Find X5 Lite & Dimensity 900   & 15 & \cmark & \xmark & \ding{109} & \xmark & \ding{109} & \xmark & \ding{109} \\
      iPhone 13 Pro     & Snapdragon X60  & 15  & \cmark & \xmark & \ding{109} & \xmark & \ding{109} & \xmark & \ding{109} \\
      Google Pixel 8    & Exynos 5300i    & 16  & \cmark & \ding{108}    & \ding{109} & \ding{108}  & \ding{109} & \xmark & \ding{109} \\
      Galaxy S23        & Snapdragon X70  & 16  & \cmark & \LEFTcircle & \ding{109} & \ding{108}  & \ding{109} & \xmark & \ding{109} \\
      Quectel RM520N-GL & Snapdragon X62  & 16  & \cmark & \ding{108}    & \ding{109} & \ding{108}  & \ding{109} & \xmark & \ding{109} \\
      
      \bottomrule
      \addlinespace[2mm]
      \multicolumn{10}{l}{\footnotesize \ding{108}: Concealed SUCI \quad \ding{109}: Clear-Text IMSI \quad \LEFTcircle: Clear-Text IMSI 2010 SIM Card \quad \xmark: 5G Core not available for the subscriber} \\
    \end{tabular}
  }
\end{table*}

\noindent{\textbf{Results:}}
The results of the experiments performed in SA commercial networks are also depicted in Table~\ref{tab:imsi_attack} using black and white circles depending on the collected data (SUCI/Clear-Text IMSI). 

In our evaluation, three out of seven devices connected to the 5G SA networks of Operator A and Operator B successfully implemented identity concealment as expected. However, a notable exception was observed with the Galaxy S23 on Operator A's network. When using a specific legacy SIM card, the terminal failed to generate a SUCI and instead transmitted the IMSI in cleartext. 

The remaining devices were not able to connect to 5G SA cells even when modem manufacturers claimed compatibility. The situation is further exacerbated in the case of \textit{Zflip 3}, \textit{Oppo Find X5 Lite and iPhone 13 Pro}, which were able to establish a connection with our laboratory 5G SA cell but were not capable of establishing connectivity with commercial ones for any of the studied operators. The findings showcase that the operation bands of the selected operators were not supported by the terminals and hence, no SA connectivity was available for them to attach. This limitation not only affects the users in terms of services but also in the security mechanisms available for them.

\begin{figure*}[t!] 
    \centering
    \includegraphics[width=0.9\textwidth, alt={Setup image}]{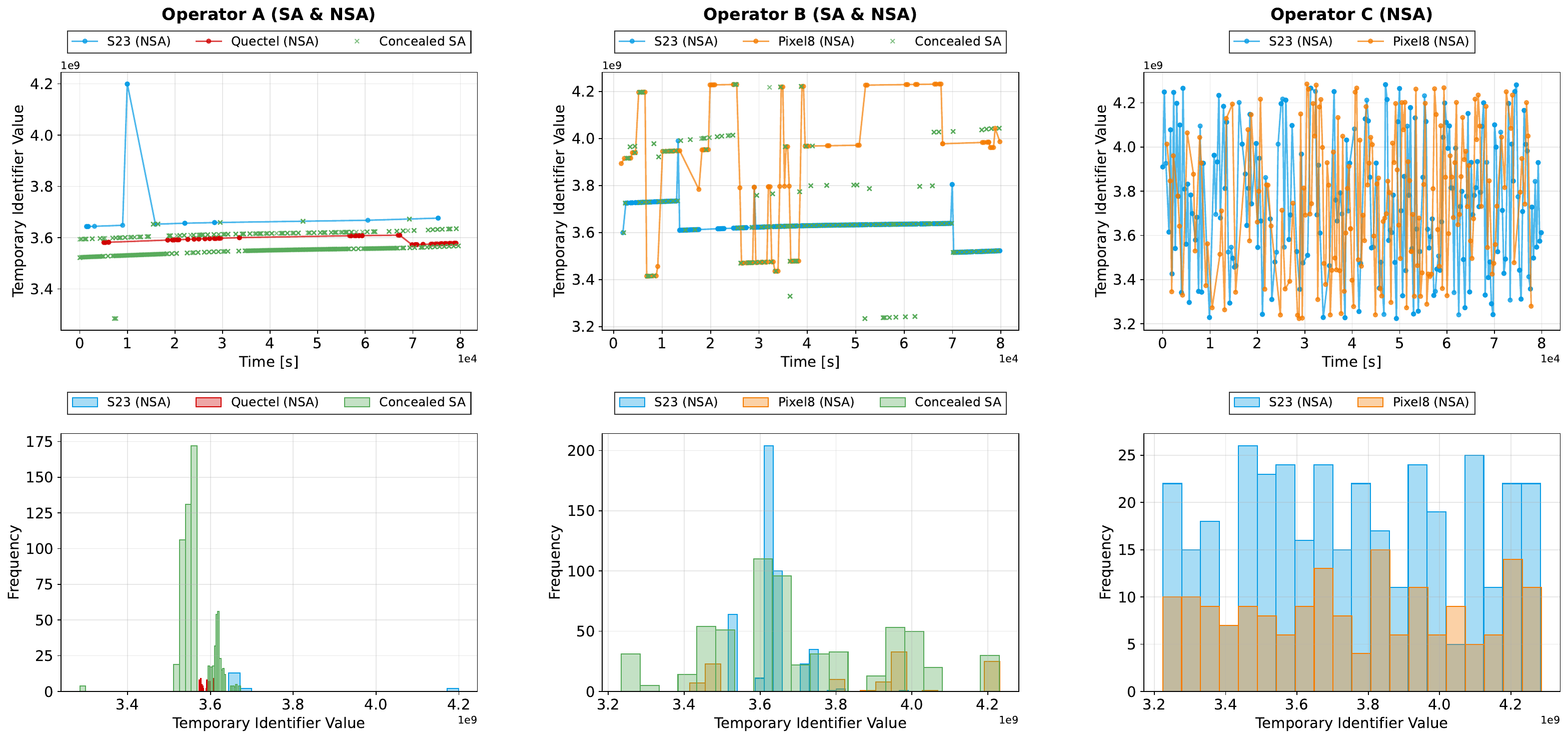}
    \caption{Comparative analysis of temporary identifier allocation strategies. Each column corresponds to one operator (A, B and C), the top row shows the temporal evolution of assigned GUTIs, and the bottom row shows the corresponding frequency histograms.}
    \label{fig:guti-eval}
    \vspace{-4mm}
\end{figure*}

\hfill

\noindent\textbf{Retry burst and cyclic backoff} 
\hfill

The successful execution of the IMSI Catching attack allowed us to perform further experiments, where we found two interesting behaviours:
\begin{itemize}
    \item \textbf{Targeted Rejection and Retry Burst:} We configured the core network of the \newtool{} entity to send a \textbf{5GMM Cause \#91} (Data Network Name (DNN) not supported) when replying to the \texttt{Registration Request}. UE-side traces reveal that the device did not accept this rejection immediately, but, instead, it entered a rapid retransmission burst, repeating the \textit{Registration Request} sequence (\textit{ID Leakage}) exactly \textit{five times} before ceasing communication with the \newtool{} cell.
    
    \item \textbf{Cyclic Back-off and Entrapment:} Following the failure previously described, the subscriber entered a back-off state governed by the standard \textbf{T3502 timer} (12 minutes), where the subscriber fell back to the legacy 5G NSA legitimate network. Upon \textit{T3502 timer} expiration, the subscriber automatically reinitiated registration with \newtool{}. This confirms that once the loop is established, the subscriber remains entrapped in a perpetual cycle.
\end{itemize}

\subsubsection{\textbf{Subscriber traceability audit in commercial scenarios}}
\label{subsec:guti-reloc}

Subscriber traceability largely depends on avoiding any link between temporary identifiers and any other type of information that may uniquely identify a subscriber. Hence, auditing how networks generate temporary identifiers and the refresh frequency is fundamental to maintaining the user's privacy. The raw records collected by \newtool{} are reduced to one record per registration in the form of a dataset storing timestamp, operator, device identifier, technology, permanent identity used and GUTI. In total, we collected $3742$ GUTI observations across NSA (4G GUTI and IMSI) and SA (5G GUTI and SUCI). The identifiers are captured as the network assigns them under repeated registrations, and hence, the dataset is a longitudinal view of each operator's temporary-identifier allocation behaviour.

\noindent{\textbf{Attack Execution Flow:}}
The adversary will follow a similar configuration and steps as described in Sections~\ref{subsec:wild_nsa} and \ref{subsec:wild_sa}, through which a subscriber will be forced to perform multiple \texttt{Registration Request} or \texttt{Mobility Request} procedures during a certain period of time through the influence of \newtool{}.
By accepting the request of each registration attempt and subsequently releasing the connection (or triggering a re-authentication challenge), \newtool{} will force the subscriber to establish a new context with the legitimate network. This process involves the assignment of a new 5G-GUTI from the core network (generated by the AMF) that will later on be collected by the adversary through \newtool{}. These experiments were performed using multiple SIM cards from the Operators under study to enrich the collected dataset. 

\hfill

\noindent{\textbf{Results:}} The experimental results regarding temporary identifier generation are presented in Fig.~\ref{fig:guti-eval}, categorised by the three studied operators (A, B, and C). The upper row illustrates the temporal evolution of the generated temporary identifiers (GUTIs) assigned to individual subscribers, while the bottom row displays the corresponding frequency histograms of these collected identifiers. To facilitate visual evaluation, distinct colours are assigned to identifiers belonging to specific subscribers.

These datasets were constructed by extracting the GUTI from the initial \texttt{REGISTRATION\_REQUEST} sent to \newtool{} and mapping it to the identity provided in the subsequent \texttt{IDENTITY\_RESPONSE} within the same registration burst. In the 5G SA networks (Operators A and B), the permanent identifier is concealed (SUCI), rendering traditional IMSI catching insufficient for direct user identification. 

The temporal evolution analysis reveals a stark contrast in operator security policies. Operator C demonstrates a robust identifier generation strategy, issuing values that are randomly and uniformly distributed regardless of the target subscriber. Conversely, Operator A (both technologies) and Operator B under NSA exhibit highly predictable, traceable patterns over time where each subscriber's GUTI seems to advance in small, near-sequential increments, keeping successive values tightly clustered and linkable across registrations.

The spatial distribution data further corroborates this vulnerability. While operator C spreads consecutive GUTIs uniformly across its identifier range, Operator A (both technologies) and Operator B under NSA keep each subscriber within a narrow, slowly advancing value band. These per-subscriber bands map directly to individual devices, and their small inter-registration step is what makes persistent user tracking feasible.

Operator B SA, however, does not clearly show in the figure a trend that reveals a predictable pattern on GUTI reallocation. Hence, we further analyse the temporary-identifier allocation by studying the \emph{step size}, computed as the median $|\Delta\text{GUTI}|$ between a subscriber's consecutive re-registrations:

\begin{equation}
\text{Step Size} = \frac{\text{median}\left( \left| \text{GUTI}_{i,t} - \text{GUTI}_{i,t-1} \right| \right)}{2^{32}} \times 100\%
\end{equation}

A step is the $|\Delta\text{GUTI}|$ between two consecutive registrations of the same subscriber stream counting only reallocations ($\Delta\text{GUTI} \neq 0$), defining a subscriber stream as the one subscriber's GUTIs ordered by time. In the case of NSA grouped by the retrieved IMSI and in SA, where the identity is concealed, streams are reconstructed using greedy nearest-value assignment (a new track is open only when a value is $> 1\%$ of $2^{32}$). All per-stream steps in a group are pooled before the median is taken so that jumps between different subscribers are not counted as steps.

Let group's GUTI values be $g_1, g_2, ..., g_n$, $span = \max(g_i) - \min(g_i)$, and let $S = $ \{pooled non-zero steps\}. To make allocators comparable we normalise the step to each one's \emph{effective range}. Every operator in our dataset fixes the two most significant bits of the identifier, so GUTIs occupy only $16$--$25\%$ of the nominal $2^{32}$ space (column ``Eff.\ range'' in Table~\ref{tab:step_size}). Expressing the step (median) as a fraction of that used span allows reading it directly against the uniform re-randomiser expectation, whose median step is $\approx29\%$ of the range. We complement the step with the \emph{linkable} share, the fraction of a subscriber's consecutive re-registrations that a nearest-value linker keeps within $1\%$ of $2^{32}$.

\begin{table*}[t]
\centering
\caption{Temporary-identifier traceability per operator and RAT. \textbf{Eff. range}: used value span as \% of $2^{32}$ (below $100\%$ as top bits are operator-fixed). \textbf{Step}: median $|\Delta\text{GUTI}|$ between a subscriber's consecutive re-registrations, as \% of $2^{32}$ and of the effective range (uniform re-randomiser $\approx29\%$; $\to0\%$ is quasi-sequential). \textbf{Linkable}: share of consecutive re-registrations a nearest-value linker keeps within $1\%$ of $2^{32}$. Steps are pooled per subscriber.}
\label{tab:step_size}
\begin{tabular*}{\textwidth}{@{\extracolsep{\fill}}l c c c c c l@{}}
\toprule
\textbf{Operator / Tech} & \textbf{Records} & \textbf{Eff.\ range} & \textbf{Step} & \textbf{Step} & \textbf{Linkable} & \textbf{Reading} \\
                         &                   & (\% $2^{32}$)        & (\% $2^{32}$) & (\% range)    &                   &                   \\
\midrule
Operator A / SA  & 1422 & 17.2\% & 0.006\% & 0.04\%  & 96\%        & quasi-sequential \\
Operator A / NSA & 130  & 16.5\% & 0.018\% & 0.11\%  & 84\%        & quasi-sequential \\
Operator B / SA  & 628  & 23.2\% & 0.636\%  & 2.74\%  & $\sim$50\%  & partially re-randomised \\
Operator B / NSA & 564  & 19.2\% & 0.005\% & 0.03\%  & 90\%        & quasi-sequential \\
Operator C / NSA & 998  & 24.7\% & 7.18\%  & 29.05\%  & 7\%         & fully re-randomised \\
\bottomrule
\end{tabular*}
\end{table*}
 
Table~\ref{tab:step_size} reads directly as a traceability ranking. Operator~A under both radio access technologies and Operator~B under NSA advance in quasi-sequential steps of at most $0.11\%$ of their effective range, two orders of magnitude below the $\approx29\%$ uniform expectation, so $84$--$96\%$ of a subscriber's consecutive re-registrations remain value-linkable and the subscriber is trivially trackable. Operator~C/NSA sits at the opposite extreme: its median step is $28.9\%$ of the range, essentially the uniform ideal, and only $7\%$ of consecutive re-registrations stay linkable, so value-based tracking collapses. Operator~B/SA is intermediate: about half of consecutive re-registrations remain linkable, so its re-randomisation degrades but does not defeat tracking completely. This dependence on both operator and
radio access technology indicates that the residual exposure is an implementation choice rather than a property mandated by the standard.

This predictable, quasi-sequential allocation is categorised as an \texttt{implementation gap}: the generation of temporary identifiers is implementation-dependent, and a small inter-registration step lets an observer link successive GUTIs and, through the same-burst mapping, tie them to the concealed SUCI, enabling user identification.

Moreover, an interesting finding in the collected traces is that the devices used for this study showed a pattern in the reallocation rate (fraction of registrations that get a new GUTI) that is consistent across NSA operators. For example, Pixel 8 shows a reallocation rate of $\sim90\%$ while Samsung Galaxy S23 keeps the rate at $\sim55\%$. This evidence may decouple the fingerprint from the time confound, given that until now, we have focused our study on the range and step magnitudes of the GUTI, which are given by the network. Unfortunately, these findings are not present for SA, disclosing a more uniform GUTI refresh policy in the 5G Core Network when compared with the 4G Core.

\subsection{Unauthenticated Signalling Resilience}
\label{subsec:RejectCause}

We further study the behaviour of subscribers when they receive unauthenticated signalling messages. More precisely, we analyse how devices operating on two real-world operators react upon receiving different \textit{Registration Reject} codes when camping on \newtool{} to replicate downgrade and DoS attacks similar to the ones described in~\cite{shaik2015practical,kim2019touching}.

Table \ref{tab:reg_reject_comparison} summarises the rejection causes evaluated in this study. The first column lists the specific rejection cause codes as they are defined by 3GPP. The second column categorises these causes according to the taxonomy defined in 3GPP TS 24.501, Annexes A and B~\cite{ts24501}. 3GPP groups these failures into four primary domains: i) causes related to subscriber Identification (subscriber ID), ii) Subscription options, iii) PLMN- or SNPN-specific network failures (Network PLMN), and iv) Invalid Messages. The following two columns detail the experimental results obtained using a Samsung Galaxy S23 terminal across two operators providing SA coverage (Operators A and B). Then, the \textit{Vulnerability Category} column indicates the impact of these codes from the subscriber perspective. Finally, the last two columns indicate whether the identified behaviour is a design limitation, meaning that it is a behaviour defined within the standard, or an implementation gap, which means that implementation decisions on either the modem or the network lead to the specific behavioural result.

\begin{table*}[t]
    \centering
    \small
    \renewcommand{\arraystretch}{1.4} 
    \rowcolors{2}{gray!10}{white}    
    
    \caption{Subscriber behaviours upon reception of different Registration Reject codes while connected to two different 5G SA Operators: Operator A vs. Operator B}
    \label{tab:reg_reject_comparison}
    
    \begin{tabular}{l c c c c c c}
        \toprule
        \textbf{3GPP}           & \textbf{3GPP}     & \textbf{Operator}     & \textbf{Operator}  & \textbf{Vulnerability}    & \textbf{Design}   & \textbf{Impl.}    \\
        \textbf{Reject Codes}   & \textbf{Category} & \textbf{A (SA)}       & \textbf{B (SA)}    & \textbf{Category}         & \textbf{Limit.}   & \textbf{Gap}      \\
        \midrule
        \rowcolor{red!25}
        \textbf{3, 6}       & UE ID             & UMTS (No Service)                             & UMTS (No Service)                             & Persistent Loss   & \cmark    &           \\
        \textbf{9, 10}      & UE ID             & 5x Requests $\rightarrow$ LTE (5G NSA)        & 5x Requests $\rightarrow$ LTE (5G NSA)        & Downgrade         & \cmark    &           \\
        \rowcolor{orange!25}
        \textbf{11}         & Subscription      & \textbf{Infinite Loop}                        & \textbf{Infinite Loop}                        & DoS               &           & \cmark    \\
        \rowcolor{yellow!25}
        \textbf{12}         & Subscription      & \textbf{Frozen Modem}                         & \textbf{Frozen Modem}                         & DoS               &           & \cmark    \\
        \textbf{13}         & Subscription      & UMTS $\rightarrow$ LTE $\rightarrow$ 5G SA    & UMTS $\rightarrow$ LTE $\rightarrow$ 5G SA    & Downgrade         & \cmark    &           \\
        \textbf{15}         & Subscription      & 5G SA (Different NR Band)                     & 5G SA (Different NR Band)                     & None              &           &           \\
        \textbf{5, 72}      & Subscription      & 5x Requests $\rightarrow$ LTE (5G NSA)        & 5x Requests $\rightarrow$ LTE (5G NSA)        & Downgrade         & \cmark    &           \\
        \textbf{27}         & Subscription      & LTE (5G NSA)                                  & LTE (5G NSA)                                  & Downgrade         & \cmark    &           \\
        \rowcolor{red!25}
        \textbf{7}          & Subscription      & UMTS (No Service)                             & UMTS (No Service)                             & Persistent Loss   &           & \cmark    \\
        \begin{tabular}{@{}l@{}} 
            \textbf{20--24, 26, 28, 43,} \\ 
            \textbf{65, 67, 69, 71,} \\ 
            \textbf{90--92} 
        \end{tabular}       & Network PLMN      & 5x Requests $\rightarrow$ LTE (5G NSA)        & 5x Requests $\rightarrow$ LTE (5G NSA)        & Downgrade         & \cmark    &           \\
        \rowcolor{orange!25}
        \textbf{73}         & Network PLMN      & \textbf{Infinite Loop}                        & \textbf{Infinite Loop}                        & DoS               &           & \cmark    \\
         
        \textbf{95--97, 99, 111} & Invalid Msgs & LTE (5G NSA)                                  & LTE (5G NSA)                                  & Downgrade         & \cmark    &           \\
        \textbf{98, 100, 101}    & Invalid Msgs & 5x Requests $\rightarrow$ LTE (5G NSA)        & 5x Requests $\rightarrow$ LTE (5G NSA)        & Downgrade         & \cmark    &           \\
         
        \bottomrule
    \end{tabular}

\fbox{
  \small
  \quad
  \legendbox{gray!50!black}{gray!20} \hspace{3pt} Soft Rejections \quad \quad
  \legendbox{red}{red!25} \hspace{3pt} Hard Rejections \quad \quad
  \legendbox{orange}{orange!25} \hspace{3pt} Infinite Loop \quad \quad
  \legendbox{yellow!90!black}{yellow!25} \hspace{3pt} Frozen Modem State \quad
}
\end{table*}

\noindent\textbf{Results:}
Based on the experimental results in Table~\ref{tab:reg_reject_comparison}, we observe two distinct redirection strategies depending on the severity of the rejection:

\begin{itemize}
    \item LTE/NSA Retention (Soft Rejections): For most mobility and congestion-related codes (e.g., \#9, \#10, \#13, \#5, \#72, \#27, \#20--\#92, \#95--\#111 and \#98--\#101), both operators implement an \textit{LTE Retention} strategy. The subscriber is redirected to LTE (5G NSA), maintaining service continuity, but privacy features like SUCI concealment are unavailable. 
    
    \item Deep Fallback (Hard Rejections): For identity-related rejections (\#3, \#6) or regional bans for 5G services (\#7), both operators force a \textit{Deep Fallback} to legacy infrastructure. In these cases, the subscriber is pushed down to UMTS in a ``Limited Service'' state. As indicated in Table~\ref{tab:reg_reject_comparison}, this results in no data service being available to the user. The 5GS capability is disabled, and the terminal is restricted to the minimum signalling required for emergency calls, significantly degrading the user's connectivity. \#3, \#6 are clearly \texttt{design limitations}, but in \#7, the subscriber should retain non-5GS/EPS (LTE) service still available, which discloses an \texttt{implementation gap} also present.
\end{itemize}

In the majority of tested scenarios, both operators exhibit completely similar behaviour, which is compliant with 3GPP specifications (\texttt{design limitation}). For instance, upon receiving \textbf{Cause \#3} (Illegal subscriber) or \textbf{\#6} (Illegal ME), the device correctly transitions to a ``Limited Service'' state on UMTS. As dictated by the standard, the subscriber considers the USIM invalid for 5GS/EPS services and ceases further access attempts to the current cell. Other compliant behaviours observed include successful fallback to 4G LTE/5G NSA (e.g. \textbf{Cause \#5, \#72, ...}) or the autonomous selection of alternative 5G SA cells when the current Tracking Area or PLMN is restricted (e.g. \textbf{Cause \#15}).

Despite the standardised directives, the subscriber exhibited critical implementation-specific failures when faced with specific subscription and network rejection codes categorised within the \texttt{Implementation Gap}: 
\begin{itemize}
    \item Infinite Loop (\#11 \& \#73): Upon receiving \textit{PLMN not allowed} (\#11) or \textit{Serving Network not authorized} (\#73), the subscriber enters a continuous, high-frequency retry loop regardless of the operator. Per TS 24.501, the subscriber should either add the PLMN to the ``Forbidden PLMN'' list or adhere to the \textit{T3346} back-off timer. The observed behaviour suggests a failure in the NAS layer to honour these timers or update the forbidden list, resulting in a signalling-based battery exhaustion attack.
    
    \item Frozen Modem State (\#12): When encountering the \textit{Tracking Area not allowed} code, the standard procedure requires the subscriber to perform a new cell selection. Instead, the Samsung Galaxy S23 terminal became completely unresponsive (modem crash), requiring a network reboot (airplane mode ON/OFF). This indicates a critical implementation flaw in the terminal's modem firmware when handling specific NAS signalling transitions.
\end{itemize}

\noindent{\textbf{Security Takeaways}:} The systematic injection of various \textit{Registration Reject} codes provides a granular assessment of how security policies are enforced across the network and the subscriber. Our experimental results demonstrate that specific rejections do not merely terminate a connection; instead, they trigger unhandled states that expose the user to secondary security vulnerabilities and operational failures.

First, a critical privacy risk is observed when the subscriber is forced into a \textit{Deep Fallback} to legacy infrastructure. While 5G provides end-to-end identity protection via SUCI, a downgrade to 4G LTE or UMTS degrades this protection. For rejections that force the subscriber into UMTS (e.g., Causes \#3, \#6, and \#7), the security and operational risks are maximised. This ``Limited Service'' state functions as a de facto DoS, as it strips the device of all data connectivity and restricts it to emergency signalling only. This behaviour effectively bypasses 5G privacy enhancements, leaving the user vulnerable to legacy IMSI Catcher and location tracking.

Furthermore, the Infinite Loop behaviour (Causes \#11 and \#73) and the Frozen Modem State (Cause \#12) represent severe implementation-level vulnerabilities. These anomalous behaviours cause unnecessary signalling storms and excessive battery drain, effectively executing a logic-based DoS attack that renders the UE unusable until a network reboot is performed.

\section{Mitigation Strategies}
\label{sec:defense_intuition}

The transition from \texttt{RRC\_IDLE} to \texttt{RRC\_CONNECTED} represents the most critical vulnerability window in both 5G NSA and SA networks. Because the cell reselection process relies entirely on unauthenticated broadcast messages, an adversary can coerce devices into camping on malicious nodes by manipulating logical parameters (e.g., \textit{Cell Reselection Priority}) rather than relying solely on physical signal amplification.

The primary defence mechanism standardised by 3GPP is the SUCI. By forcing the encryption of the permanent identifier (SUPI) using the home network's public key before over-the-air transmission, 5G aims to render the capture of initial registration (pre-authentication) messages useless to an attacker. In theory, this shifts the decryption responsibility to the Unified Data Management (UDM) in the core network, protecting the user even if they camp on an F-BS. 

Consequently, a robust defence strategy must move beyond cryptography and incorporate subscriber-side anomaly detection. Since the F-BS relies on \textit{Absolute Priority} to force reselection, the subscriber should analyse the trustworthiness of broadcast parameters. For instance, the sudden appearance of a new cell claiming high priority, which was not advertised in the neighbour list of the previous legitimate serving cell, constitutes a logical inconsistency. By cross-referencing observed SIB parameters with historical data or neighbour relations, the subscriber could flag such cells as suspicious before initiating the vulnerable registration procedure.

Then, defences must also address the manipulation of unauthenticated Reject Causes. Current modems accept \texttt{REGISTRATION REJECT} messages even if the network has not yet authenticated itself. A heuristic defence could involve a \textit{Quarantine State}, where the subscriber treats unauthenticated rejection causes that force a downgrade (Inter-RAT transition) with scepticism, potentially reattempting connection to other 5G cells before accepting the downgrade to a less secure generation.

Finally, we propose the implementation of a certain logical consistency audit in the UE so that the actual user is aware of situations that may risk her/his privacy, e.g., implementing a visual warning when a network requests the identity in clear text. This user-oriented feedback will provide the user with an active role in their own security and build a second security line, while infrastructure-side security may include vulnerabilities.

\section{CONCLUSION} \label{sec:conclusion}

In this work, we presented \newtool{}, a network security-posture auditor that turns the standardised cell-reselection procedure into a low-disruption vantage point from which the privacy, traceability, and resilience of commercial 5G subscribers under certain adversarial conditions can be exercised and, crucially, \emph{attributed}. Rather than pursuing novel exploits, our contribution is methodological: using only open-source stacks and inexpensive SDR hardware, \newtool{} induces a target UE onto a rogue cell without active jamming or malformed-packet injection, and classifies each resulting exposure as either a \texttt{protocol-design limitation}, exploitable even against a fully specification-compliant deployment or an \texttt{implementation gap} arising from operator or modem implementation choices. This separation is what distinguishes real, actionable risk from behaviour inherent to the standard itself.

Our audit shows that the most damaging exposures are \emph{not} attributable to a single culprit but are distributed across both axes. On the design side, the UE's obligation to present an identity on request, the processing of unauthenticated \texttt{IDENTITY\_REQUEST} and \texttt{REGISTRATION\_REJECT} messages before any security context exists, and the reversion to a cleartext 4G control plane under NSA are all standard-compliant behaviours that \newtool{} exploits. The majority of the observed \textit{Registration Reject} outcomes, including the RAT downgrades, likewise conform to TS~24.501 and are therefore design limitations that no operator can configure away. On the implementation side, we find genuine deployment gaps, such as Operator A (under both SA and NSA) and Operator B (under NSA) allocate the temporary identifier in quasi-sequential steps far below the re-randomisation expectations enabling persistent tracking despite the correct SUCI concealment. On the contrary, Operator C (NSA) re-randomises and defeats value-based tracking, and Operator B (SA) re-randomises only partially (half of the consecutive re-registrations remain linkable). Moreover, specific unauthenticated reject codes drive the modem into states with no basis in the standard, an infinite signalling loop (causes~\#11, \#73) and a frozen-modem denial of service (cause~\#12) that requires a manual radio reset. Notably, these reject-induced behaviours were consistent across both SA operators, indicating that they are determined at the modem/implementation layer rather than by operator policy.

Our audit also records measurable progress. Whereas the first wave of in-the-wild 5G measurements reported that identity concealment was frequently absent and that subscriber identifiers remained widely exposed~\cite{European_5G_Security_in_the_Wild,Fact_Checking_5G_Security,eleftherakis2024demystifying}, the SA operators we audited concealed the subscriber identity correctly in every case but one, a legacy-SIM provisioning that reverted to a cleartext IMSI. We therefore find no evidence of operator-side Null-scheme deployment at scale. The residual plaintext exposure we observed is a device/provisioning artefact rather than a network policy. This is a positive signal that deployments are closing the identity-exposure gap that dominated earlier studies, and it lets us sharpen where the real risk has migrated: not to the \emph{permanent}-identifier concealment that 5G SA now largely gets right, but to the \emph{implementation} axis (temporary-identifier predictability/step size and NAS reject-handling) where the guarantees still break.

Taken together, these results argue that 5G's privacy promises fail along two different parts that demand two different remedies. On the one hand, the design-inherent exposures can only be closed by the standards bodies, for instance, through UE-side plausibility checks on broadcast reselection priorities and a quarantine policy for unauthenticated signalling that force a downgrade. On the other hand, the implementation gaps, such as predictable, quasi-sequential GUTI allocation (small step size) and fragile reject-handling state machines, are within the immediate reach of operators and vendors and would be addressed simply by promoting privacy features to mandatory and hardening NAS error handling. To fulfil 5G's privacy promises as it powers critical infrastructure, we must clearly define standard specification fixes and local deployment fixes.

\section*{ACKNOWLEDGMENT}
This work was supported in part by the ORIGAMI Project under Grant 101139270; in part by the CERCA Programme from the Generalitat de Catalunya through the ICREA programme; and in part by the funding received from the Department de Recerca I Universitats, Generalitat de Catalunya for this project

\bibliographystyle{IEEEtran}
\bibliography{references}

\begin{IEEEbiography}
[{\includegraphics[width=1in,height=1.25in, clip,keepaspectratio]{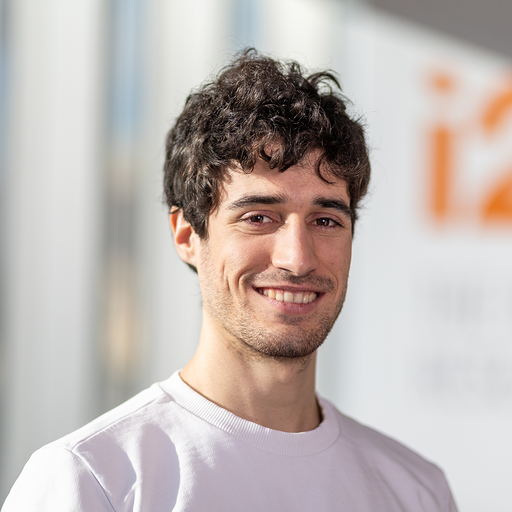}}]{Oscar Lasierra } holds a B.Eng. in Telecommunications from the Polytechnic University of Catalonia, Castelldefels. In 2022, he joined i2CAT as a junior researcher in the AI-driven Systems group. His research interests encompass a wide range of topics, including software-defined radio (SDR), cellular networks, security, and privacy.
\end{IEEEbiography}

\begin{IEEEbiography}
[{\includegraphics[width=1in,height=1.25in,clip,keepaspectratio]{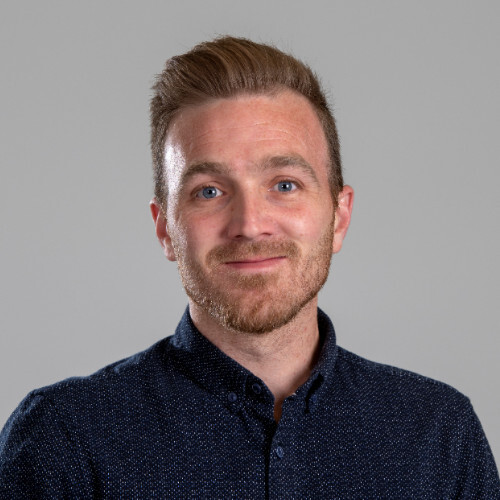}}]{Gines Garcia-Aviles } received his M.Sc. (2018) and PhD (2021, cum laude with International Mention) in Telematics Engineering from the University Carlos III of Madrid, Spain, where his dissertation received the Extraordinary Doctorate Award. During his doctoral studies, he was part of the wireless networking group at IMDEA Networks Institute and the Telematics Department at UC3M. In 2021 he joined the i2CAT Foundation as a post-doctoral researcher. From 2024 to 2026, he was a post-doctoral researcher at the University of Murcia, Spain, lecturing in mobile computing and cybersecurity. He has contributed to numerous EU-funded 5G/6G projects. His research interests include RAN virtualization and O-RAN, network slicing and orchestration, 5G/6G security, and AI-driven network management.
\end{IEEEbiography}

\begin{IEEEbiography}
[{\includegraphics[width=1in,height=1.25in,clip,keepaspectratio]{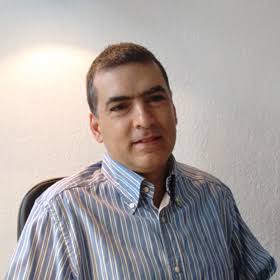}}]{Antonio Skarmeta } received a B.S. degree (Hons.), an M.S. degree in computer science from the University of Granada, Granada, Spain, and a PhD degree in computer science from the University of Murcia, Spain. He has been the Head of the Research Group, ANTS, since its creation in 1995. Since 2009, he has been a Professor at the University of Murcia. He is also an Advisor to the Vice Rector of Research with the University of Murcia, for international projects, and the Head of the International Research Project Office. 
\end{IEEEbiography}

\begin{IEEEbiography}[{\includegraphics[width=1in,height=1.25in,clip,keepaspectratio]{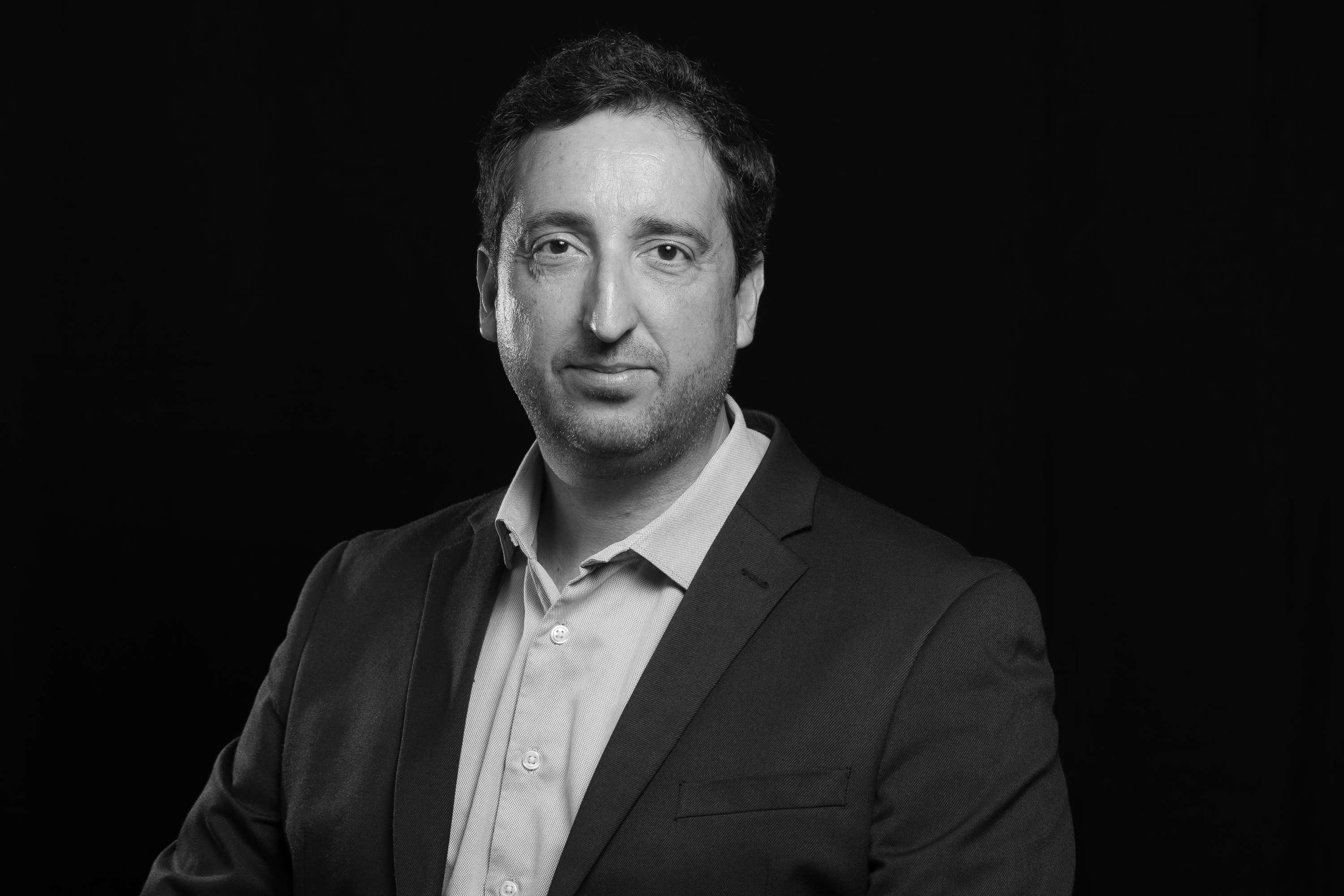}}]{Xavier Costa-Pérez }
is an ICREA Research Professor, Scientific Director at the i2CAT Research Center, and Head of 6G R\&D at NEC Laboratories Europe. He has served on the Organizing Committees of several conferences, published papers of high impact, and holds more than 80 granted patents. He received his Ph.D. degree in telecommunications from the Polytechnic University of Catalonia, Barcelona, and was the recipient of a national award for his PhD thesis.
\end{IEEEbiography}

\end{document}